\documentclass[twocolumn,preprintnumbers,amsmath,amssymb,longbibliography,aps]{revtex4}

\usepackage{graphicx}
\usepackage{dcolumn}
\usepackage{amssymb}
\usepackage{bm}
\usepackage{color}
\usepackage{enumitem}
\usepackage{mathtools}
\usepackage{comment}
\usepackage{subfigure}
\usepackage{float}

\begin{document}

\title{Divergence and Consensus in Majority Rule}

\author{P. L. Krapivsky}
\affiliation{Department of Physics, Boston University, Boston, MA, 02215 USA}
\author{S. Redner}
\affiliation{Santa Fe Institute, 1399 Hyde Park Road, Santa Fe, NM, 87501 USA}

\begin{abstract}

  We investigate majority rule dynamics in a population with two classes of
  people, each with two opinion states $\pm 1$, and with tunable interactions
  between people in different classes.  In an update, a randomly selected
  group adopts the majority opinion if all group members belong to the same
  class; if not, majority rule is applied with probability $\epsilon$.
  Consensus is achieved in a time that scales logarithmically with population
  size if $\epsilon\geq \epsilon_c=\frac{1}{9}$.  For $\epsilon <\epsilon_c$,
  the population can get trapped in a polarized state, with one class
  preferring the $+1$ state and the other preferring $-1$.  The time to
  escape this polarized state and reach consensus scales exponentially with
  population size.
\end{abstract}  
\maketitle

A major theme in the modeling of social dynamics is understanding the
conditions that cause a population to either reach consensus or a polarized
state, in which a diversity of opinions persists indefinitely (see, e.g.,
\cite{castellano2009statistical,Galam,sznajd2000opinion,galesic2019statistical}).
The voter
model~\cite{clifford1973model,holley1975ergodic,cox1989coalescing,liggett2013stochastic,krapivsky1992kinetics,frachebourg1996exact,dornic2001critical,krapivsky2010kinetic,baronchelli2018emergence,redner2019reality}
provides a simple description for consensus formation.  In a single update, a
randomly selected voter, which can be in one of two opinion states, adopts
the opinion state of a randomly selected neighbor.  Consensus is necessarily
reached in a finite population.  In contrast, polarized states arise in
models that have limited interactions between individuals of different
classes.  Prominent examples include the Axelrod
model~\cite{axelrod1997dissemination,axtell1996aligning,castellano2000nonequilibrium,vazquez2007non},
in which individuals interact only if they share a common social trait; the
bounded confidence
model~\cite{weisbuch2002meet,hegselmann2002opinion,ben2003bifurcations}, in
which individuals interact only if they are sufficiently close in opinion
space; multi-state voter models, with interactions only between voters of
compatible
states~\cite{vazquez2003constrained,marvel2012encouraging,mobilia2011fixation};
and social balance models, with edges that specify friendly or unfriendly
relations and dynamics that reduces social stress
\cite{balance05,balancePoland,balance06,balance11,balance20}.

Here, we extend majority rule dynamics
\cite{galam1999application,krapivsky2003dynamics,Mauro_Sid03,majority-PT,schoenebeck2018consensus,voter-majority,majority_networks,schoenebeck2020escaping,noonan2021dynamics}
to probe this tension between consensus and polarization in a mathematically
principled way.  The original majority rule model describes opinion evolution
in a population where each individual can be in one of two equivalent opinion
states $+1$ and $-1$.  Individual opinions change by the following steps: (i)
Pick a group of individuals from the population.  (ii) All selected
individuals adopt the opinion of the group majority.  These steps are
repeated until the population necessarily reaches consensus, either all $+1$
or all $-1$.  If individuals reside on the nodes of a complete graph, the
consensus time scales logarithmically~ \cite{krapivsky2003dynamics} with
population size---quick consensus in this mean-field limit.  For finite
dimensions, where the group consists of contiguous individuals, the consensus
time scales algebraically with population
size~\cite{chen2005majority,chen2005consensus}.

Our model, which we term the \emph{homophilous} majority rule (HMR), captures
a pervasive aspect of social interactions---namely,
homophily~\cite{mcpherson2001birds,kossinets2009origins,barbera2015birds}, in
that individuals tend to ignore the opinions of people unlike themselves.
The simplest situation is a population that consists of two classes of people
that we denote as A and B.  The update is the same as that for majority rule,
with a simple but crucial twist: (i) Pick a group of individuals at random
from the population.  (iia) If all these individuals are from the same
class, they adopt the majority opinion. (iib) If the group consists of
individuals from different classes, they adopt the majority opinion with
probability $\epsilon$; otherwise, no opinion change occurs.  The group size
should be $\geq 3$ and odd to ensure that a majority exists.  We focus on the
simplest case where each group has size 3.

When $\epsilon$ exceeds a critical value $\epsilon_c=\frac{1}{9}$, the
population quickly reaches consensus, with the average consensus time again
scaling logarithmically with population size.  When $\epsilon<\epsilon_c$,
different classes of people rarely engage and the population can get trapped
in a polarized state, with one class preferring the $+1$ state and the other
preferring the $-1$ state.  The time to escape this polarized state and
ultimately reach consensus scales exponentially with population size.  For a
macroscopic population, consensus is therefore not achieved in any reasonable
time scale.  Moreover, the distribution of consensus times contains multiple
scales, so that different instances of the population reach consensus at
wildly different times.

\smallskip\noindent\emph{Rate Equations.}  First, we treat the limiting case
of $\epsilon=1$ by the deterministic rate equation.  The density $\rho(t)$ of
individuals with opinion $+1$ evolves according
$\dot \rho = \rho^2(1-\rho)-\rho(1-\rho)^2$.  The first term accounts for the
increase in $\rho$ due to groups that consist of two individuals with opinion
$+1$ and one individual with opinion $-1$.  A parallel explanation accounts
for the second term.  The rate equation has two stable fixed points,
$\rho=0,1$, corresponding to consensus, and an unstable fixed point
$\rho=1/2$.  The average consensus time grows logarithmically with system size \cite{krapivsky2003dynamics}.

\begin{figure*}[ht]
\centerline{\subfigure[]{\includegraphics[width=0.32\textwidth]{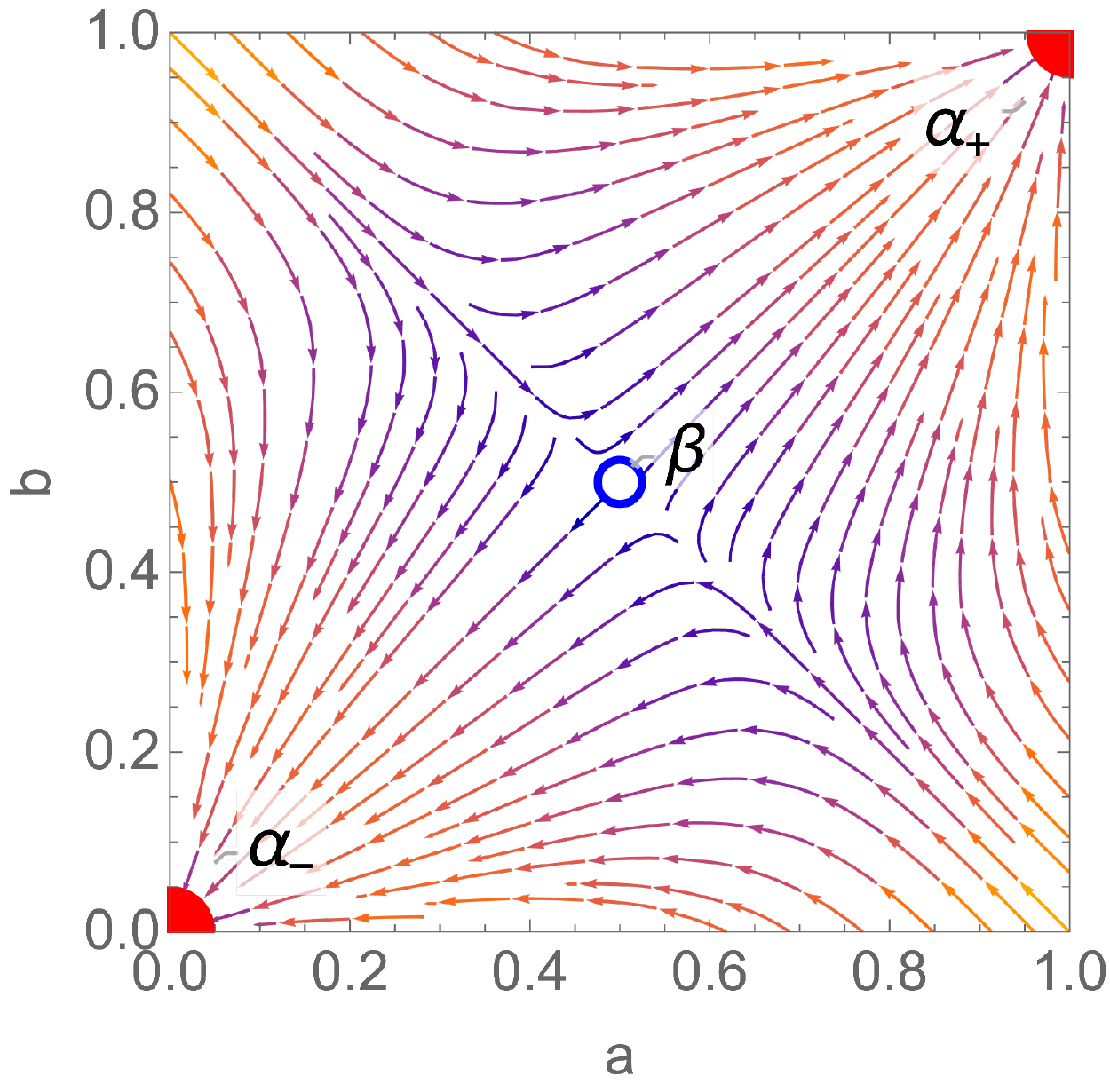}}\quad
    \subfigure[]{\includegraphics[width=0.32\textwidth]{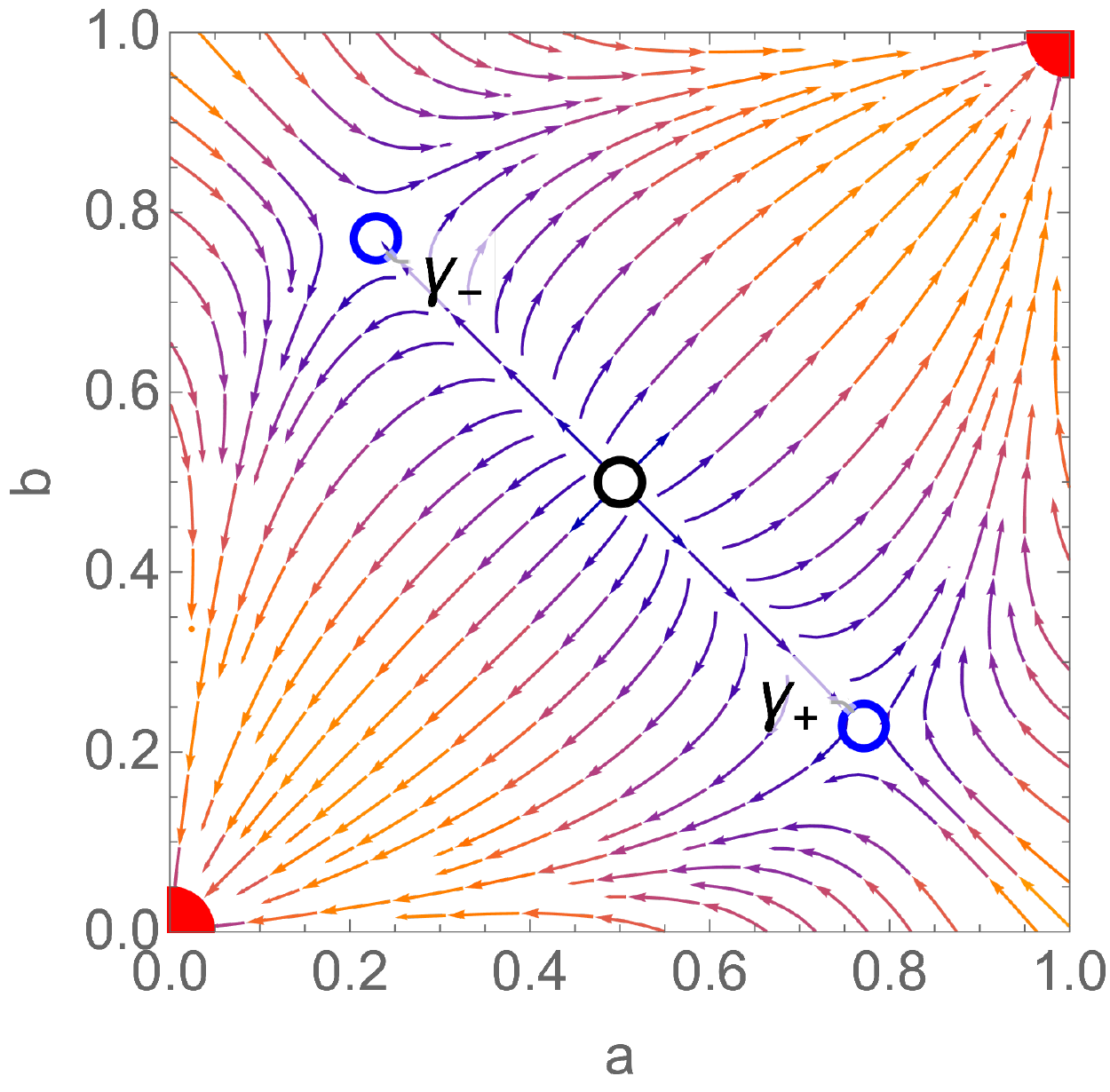}}
\subfigure[]{\includegraphics[width=0.32\textwidth]{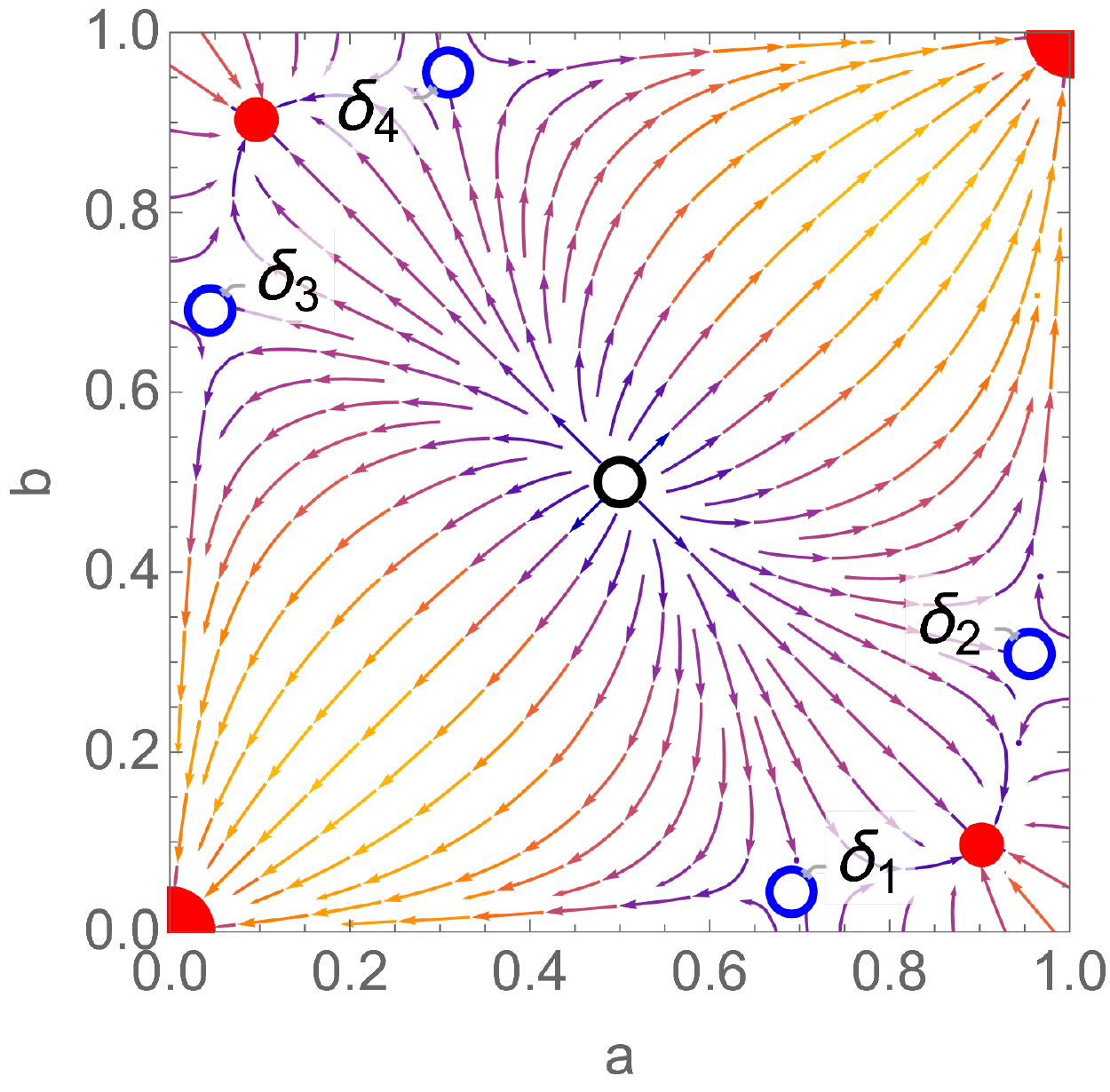}}}
\caption{Flow field of the dynamical system defined by Eq.~\eqref{RE} for the
  cases: (a) $\epsilon=0.25$ (greater than $\epsilon^c=\frac{1}{5}$), (b)
  $\epsilon=0.15$ (between $\epsilon_c=\frac{1}{9}$ and $\epsilon^c$), and
  (c) $\epsilon=0.08$ (less than $\epsilon_c$).  The arrow colors indicate
  the flow magnitude (blue slow, red fast) and the symbols indicate the
  locations of the fixed points: red dots for stable, blue circles for
  saddles, and black circle for unstable.}
\label{fig:flow}
\end{figure*}

Consider now the HMR model for arbitrary $\epsilon\leq 1$.  For simplicity,
we analyze the symmetric situation with $N$ individuals in each class.
Denote by $n_\text{A}$ and $n_\text{B}$ the number of A's and B's
with opinion state $+1$.  The densities of these two classes,
$a=n_\text{A}/N$ and $b=n_\text{B}/N$, obey
\begin{align}
\label{RE}
\dot a &= F(a) + \epsilon G(a,b)\,,\qquad \dot b = F(b)+\epsilon G(b,a)\,,
\end{align}
where 
\begin{align*}
F(x) &=   x^2(1\!-\!x)-x(1\!-\!x)^2\,, \\
G(x,y) &= (1\!-\!x)\big[2xy\!+\!y^2\big] \!-\! x\big[2(1\!-\!x)(1\!-\!y)\!+\!(1\!-\!y)^2\big]
\end{align*}
(see the supplementary material for details).  The dynamical behavior of the
system \eqref{RE} is quite rich, as illustrated by the flow field for generic
values of $\epsilon$ in each of the three domains: (i)
$\epsilon>\epsilon^c=\frac{1}{5}$; (ii) $\epsilon_c<\epsilon<\epsilon^c$ with
$\epsilon_c=\frac{1}{9}$; (iii) $\epsilon<\epsilon_c$ (Fig.~\ref{fig:flow}).

\begin{figure}[ht]
  \centerline{\includegraphics[width=0.375\textwidth]{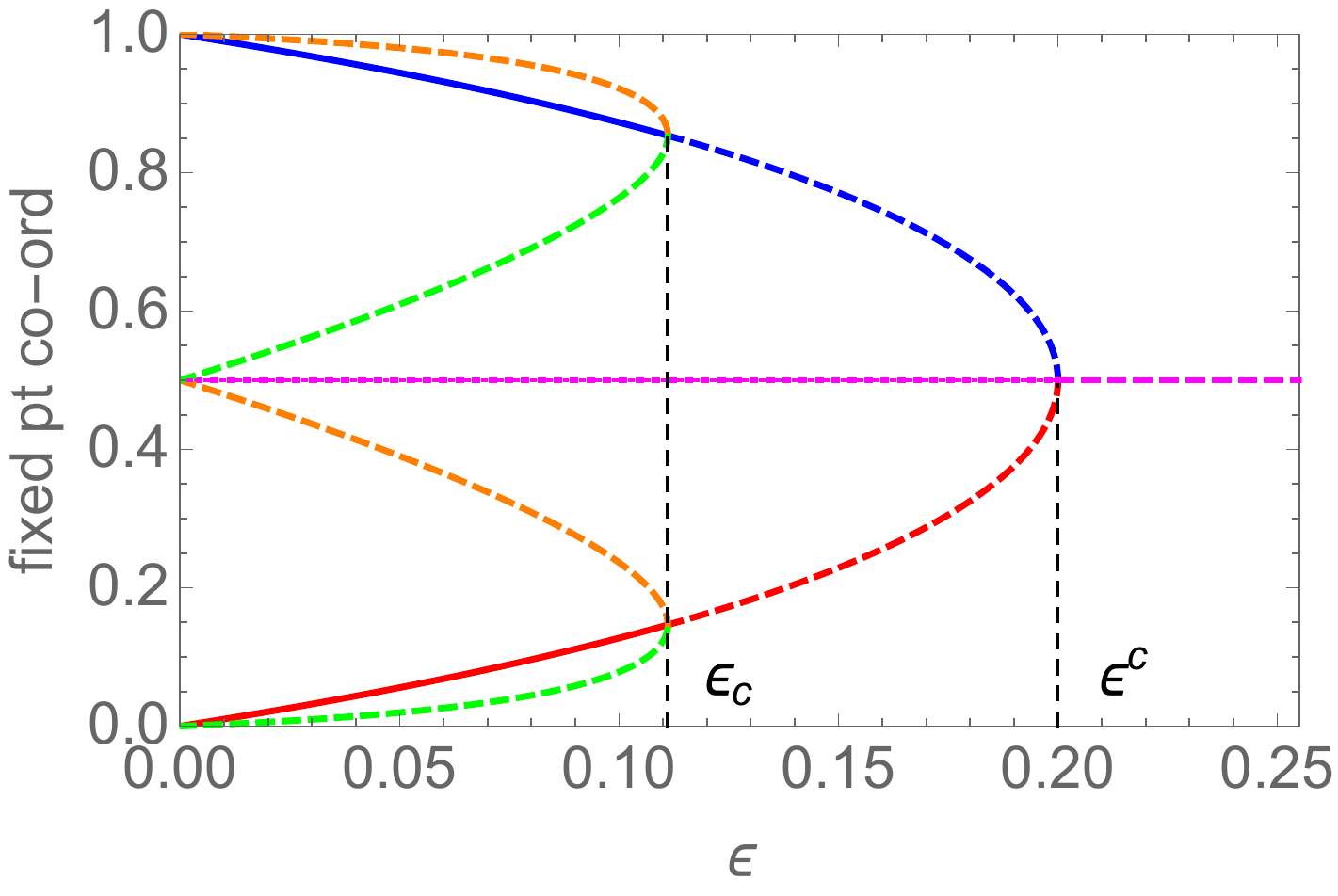}}
  \caption{The coordinates of the fixed points as function of $\epsilon$,
    with dotted, dashed, and solid indicating unstable, saddle, and stable
    nodes, respectively.  (magenta) The symmetric fixed point
    $\beta=\big(\frac{1}{2}, \frac{1}{2}\big)$.  (red/blue) The two
    reflection-symmetric fixed points $\gamma_\pm$ in the range $\epsilon=0$
    to $\epsilon^c=\frac{1}{5}$.  (green/orange) The four non-symmetric fixed
    points $\delta_i$ that emerge from $\gamma_\pm$ at
    $\epsilon_c=\frac{1}{9}$ and also extend to $\epsilon=0$.  A single color
    gives the $x$ and $y$ coordinates of a given fixed point $\delta_i$.  }
  \label{fig:fp}
\end{figure}

When $\epsilon>\epsilon^c$, the consensus fixed points at
$\alpha_-\equiv (0,0)$ and $\alpha_+\equiv (1,1)$ are stable nodes, while the
fixed point $\beta\equiv \big(\frac{1}{2},\frac{1}{2}\big)$ is a saddle.
From any initial condition that does not lie on the line $a+b=1$, the
population quickly reaches consensus at $\alpha_-$ for $a+b<1$ and consensus at $\alpha_+$
for $a+b>1$.  For initial conditions that lie on the line $a+b=1$, the
population is driven to the fixed point $\beta$.  However, stochastic
finite $N$ fluctuations drive the system from this line (and even from the
fixed point $\beta$ if the evolution begins there) and either consensus is
reached with equal probabilities.  We show below that the consensus time
scales as $\ln N$ in all cases.

\begin{figure}[ht]
  \centerline{\includegraphics[width=0.325\textwidth]{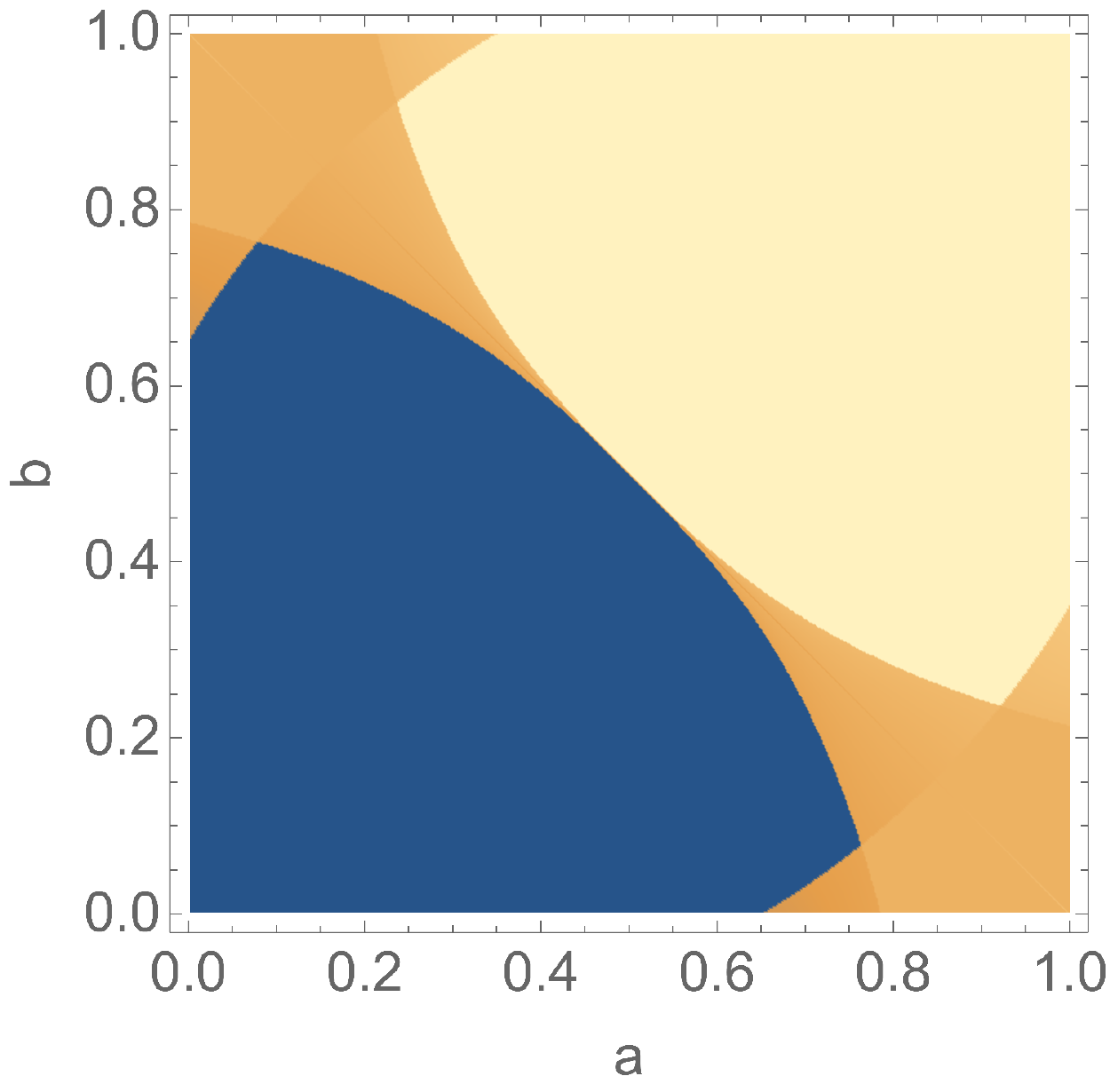}}
  \caption{The basin of attraction to the consensus fixed point (0,0) (blue)
    and (1,1) (yellow) and to the mixed-opinion fixed points $\gamma_\pm$
    (orange) for the case $\epsilon=0.1$.}
\label{fig:basin}
\end{figure}

In the intermediate regime, $\epsilon_c < \epsilon <\epsilon^c$, the fixed
point $\beta$ changes from a saddle to an unstable node and two additional
saddle-node fixed
points
\begin{equation}
\label{gamma:+-}
 \gamma_\pm=\tfrac{1}{2}\left(1\pm \Gamma\,, 1\mp \Gamma\right), \quad \Gamma=\sqrt{\frac{1-5\epsilon}{1-\epsilon}}\,,
\end{equation}
emerge from $\beta$.  These fixed points recede from $\beta$ as $\epsilon$
decreases below $\epsilon^c$ while remaining on the line $a+b=1$
(Fig.~\ref{fig:fp}).  According to the rate equations, if the initial
condition lies on the line $a+b=1$, except for $(a,b)=\beta$, the population
is drawn to one of the fixed points $\gamma_\pm$.  At $\gamma_+$, for
instance, a fraction $\frac{1}{2}(1+\Gamma)$ of A's is in the $+1$ state,
while the same majority of B's is in the $-1$ state.  Thus the total
population is polarized but evenly balanced, with one-half in the $+1$
opinion state and the other half in the $-1$ state.  All other initial
conditions are again driven to consensus.

When $\epsilon<\epsilon_c$, four additional fixed points $\delta_i$
(i=1,2,3,4) emerge, two from $\gamma_+$ and two from $\gamma_-$. These four
fixed points are saddle nodes, while the fixed points $\gamma_\pm$ become
stable.  There are now two disjoint domains in phase space
that are attractors to one of these mixed-opinion fixed points $\gamma_\pm$
(Fig.~\ref{fig:basin}).  In this regime, the population-average interaction
is sufficiently weak that A's and B's form their own and distinct
near-consensus enclaves when the initial condition is within either of these
basins of attraction for $\gamma_+$ or $\gamma_-$ (Fig.~\ref{fig:flow}(c)).
Again, the population is polarized but evenly balanced and there is range of
initial conditions for which the population is driven to this polarized
state.  Initial conditions that lie outside these two basins of attraction
are again quickly driven to one of the consensus fixed points.

\smallskip\noindent\emph{Finite-Population Simulations.}  Even for a 
perfectly-mixed population, the rate equation approach does not fully
capture the stochastic dynamics.  Because of finite-$N$ fluctuations, the only true
attractors of the dynamics are the consensus fixed points.  If the population
state is in the basin of attraction of one of the fixed points $\gamma_\pm$
(the situation pertinent for $\epsilon<\epsilon_c$), the dynamics first draws
the population to one of these fixed points.  Eventually, however, a
sufficiently large stochastic fluctuation pushes the population out of these
basins and to one of the consensus fixed points.  The probability to leave
either of these basins is exponentially small in $N$, which implies that the
time to reach consensus grows exponentially with $N$.  Even though consensus
is the true final state of the population, reaching consensus requires a time
that is practically unattainable for a population of any appreciable size.
Thus for $\epsilon<\epsilon_c$, consensus is effectively not reached.

An analogous dichotomy occurs in population dynamics models, such as the logistic process,
where the rate equation predicts a steady state, whereas extinction is the
final outcome~\cite{Karlin,Nasell,Allen}.  In these processes, extinction occurs in a time that
scales exponentially with the quasi steady-state population size predicted by
the rate equation (see,
e.g.,~\cite{elgart2004rare,kessler2007extinction,assaf2010extinction,assaf2017wkb}).
The HMR model exhibits a similar rare-event driven approach to a final
consensus, but with the additional feature that this approach is governed by
two very different time scales.

In our simulations, we first select three individuals at random from the
entire population.  If these individuals are all from the same class,
majority rule is applied.  If the group consists of different classes of
individuals, majority rule is applied with probability $\epsilon$; otherwise,
nothing happens.  The time is increment by $3/N$ in each update so that every
individual reacts once, on average, in a single time unit.  This update is
repeated until consensus is achieved.  We investigated several generic
initial conditions: (a) A fully polarized state, with A's are entirely in the
$+1$ state, and B's are entirely in the $-1$ state; (b) a ``balanced'' state,
in which half of both the A's and B's are in the $+1$ state; (c)
``imbalanced'' states, in which a fraction $q$ of the A's and a fraction
$1-q$ of the B's are in the $+1$ state.  This initial condition lies along
the line $a+b=1$ in state space.  The results for these initial conditions
are qualitatively similar and we primarily focus on the balanced initial
condition.

\begin{figure}[ht]
  \centerline{\subfigure[]{\raisebox{.25 cm}{\includegraphics[width=0.235\textwidth]{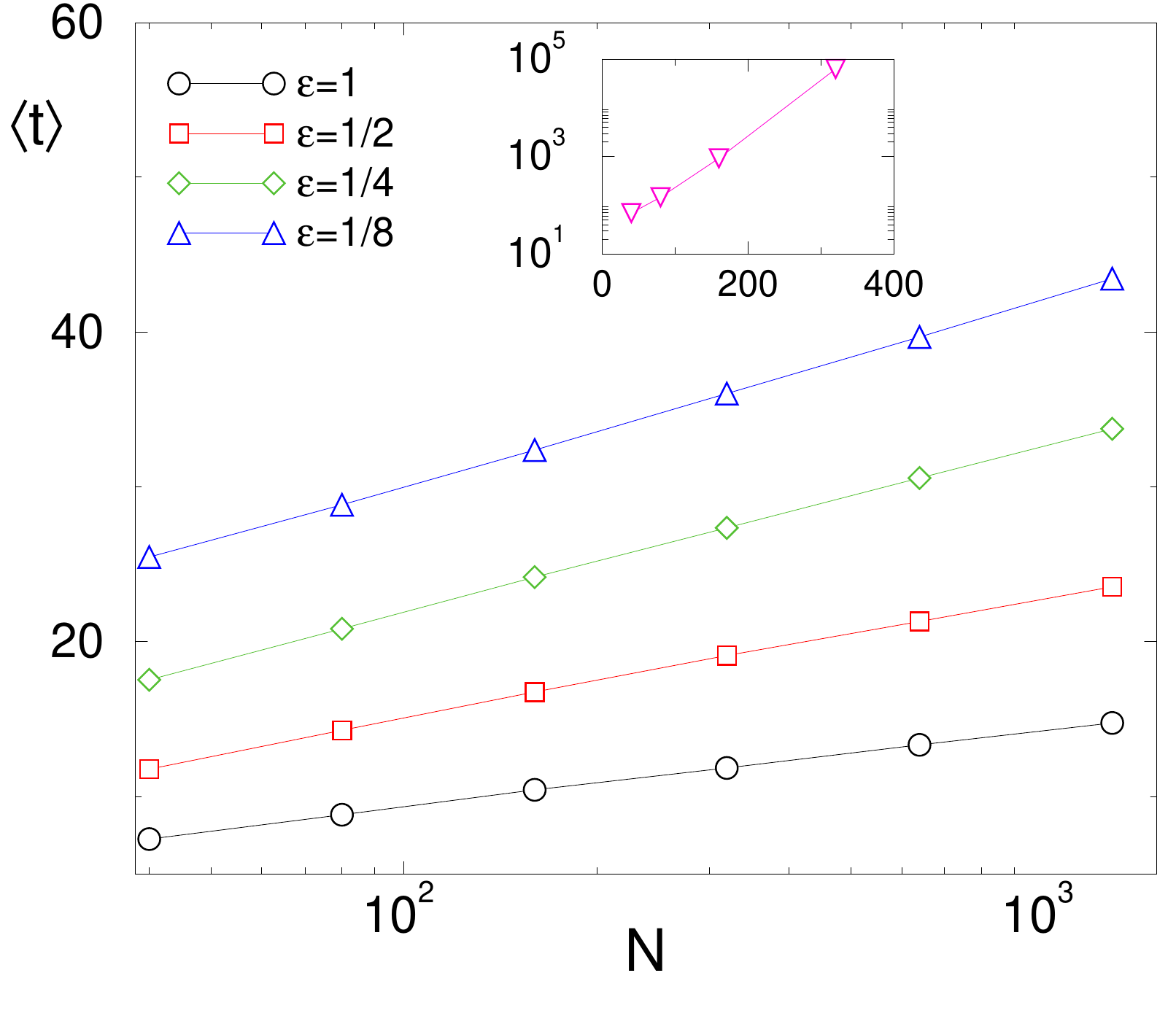}}}
    \subfigure[]{\includegraphics[width=0.245\textwidth]{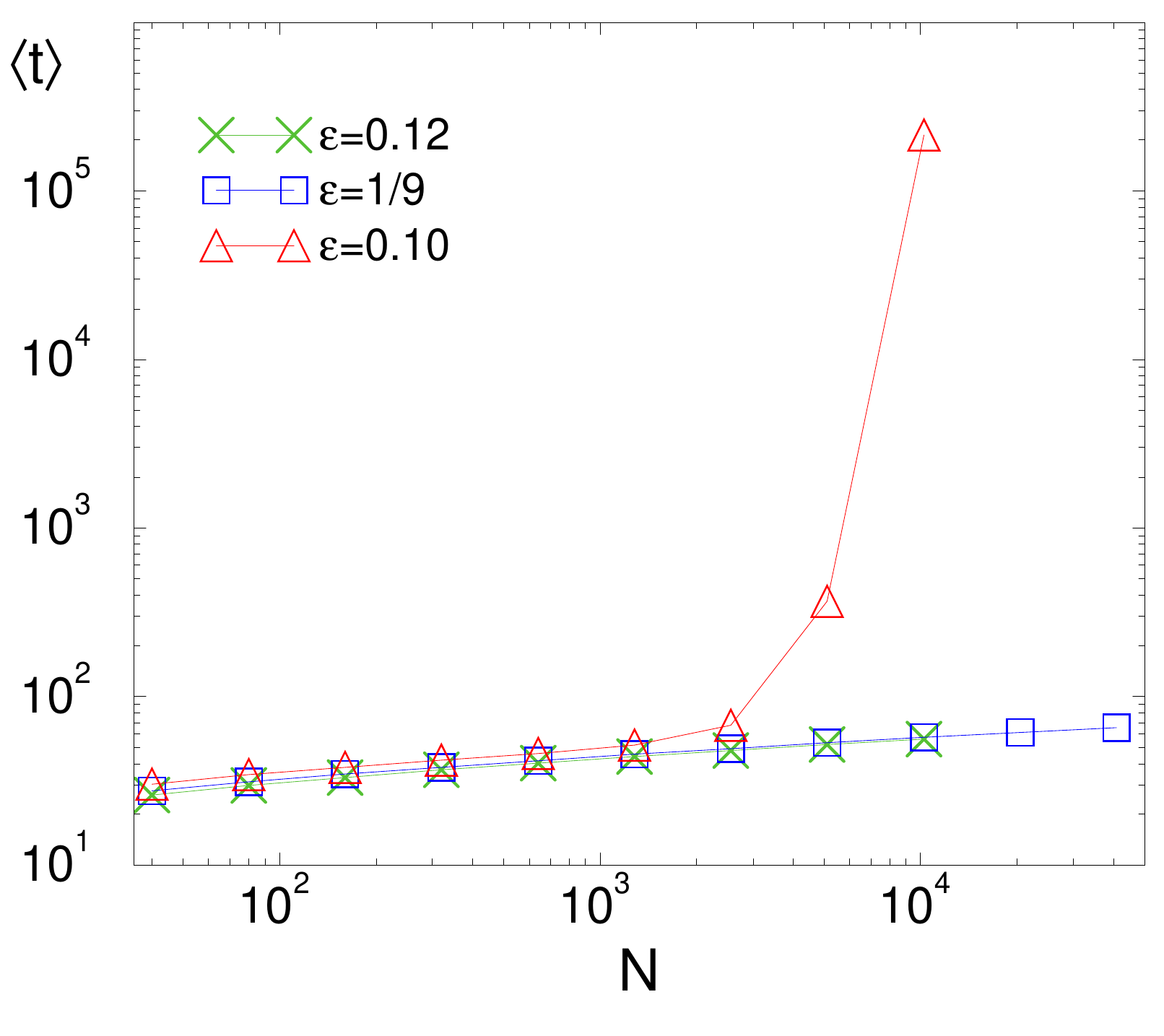}}}
  \caption{The average consensus time $\langle t\rangle$ versus $N$ for the
    balanced initial condition for: (a) a range of $\epsilon$ values and (b)
    close to the transition at $\epsilon=\epsilon_c$.  In (a), the abscissa
    is on a logarithmic scale in the main panel and the ordinate is on a
    logarithmic scale in the inset, where data for the case
    $\epsilon =\frac{1}{16}$ is shown. }
\label{fig:tav}
\end{figure}

For $\epsilon > \epsilon_c$, the average consensus time $\langle t\rangle$
grows logarithmically with $N$. This is an exact result
\cite{krapivsky2003dynamics} for the original MR model ($\epsilon=1$), and it
is also proved to occur for majority-like stochastic processes that are
described by rate equations that possess only saddles and sinks---escaping a
saddle and reaching a sink takes $O(\ln N)$
time~\cite{schoenebeck2018consensus,schoenebeck2020escaping}. This
logarithmic dependence also occurs in the regime
$\epsilon_c\leq \epsilon\leq \epsilon^c$.  In this intermediate regime, when
the initial state is on the line $a+b=1$, finite-$N$ fluctuations will drive
the population state off this line (where $\gamma_\pm$ are attractors), after
which consensus is quickly reached.

When $\epsilon$ decreases below $\epsilon_c=\frac{1}{9}$, the $N$ dependence
of $\langle t\rangle$ suddenly changes from logarithmic to exponential
(Fig.~\ref{fig:tav}(b)). Strikingly, this exponential dependence sets in only
after $N\agt 4000$ for the case $\epsilon=0.1$, a feature that arises from
from the geometry of the basin of attraction (Fig.~\ref{fig:basin}(b)).  To
reach one of the fixed points $\gamma_\pm$ when starting from
$(a,b)=(\frac{1}{2},\frac{1}{2})$, the population state has to navigate
within the tongue between the separatrices that border the basins of
attraction to $\gamma_\pm$.  This tongue is narrow when $\epsilon$ is close
to $\epsilon_c$, so the population state is typically and quickly drawn to a
consensus fixed point in this range.  However, if either fixed point
$\gamma_\pm$ is reached, the escape time scales exponentially with $N$.
These two outcomes explain the existence of two drastically different time
scales in $P(t)$, the consensus-time distribution (Fig.~\ref{fig:tdist}).  We
may estimate the probabilities of the two outcomes from the data for $P(t)$
and find that the probability to reach one of the fixed points $\gamma_\pm$
vanishes as $\epsilon\to\epsilon_c$ from below; a bound for this probability
is given in the supplemental material.  Finally, note that this bimodal
consensus-time distribution (Fig.~\ref{fig:tdist}) arises when the starting
state is $(a,b)=(\frac{1}{2},\frac{1}{2})$.  If the starting point lies
inside the basin of attraction of $\gamma_\pm$, the consensus-time
distribution is asymptotically exponential,
$P(t)=\langle t\rangle^{-1}e^{-t/\langle t\rangle}$ and is fully
characterized by the average consensus time~\cite{assaf2017wkb}.

\begin{figure}[ht]
  \centerline{\includegraphics[width=0.325\textwidth]{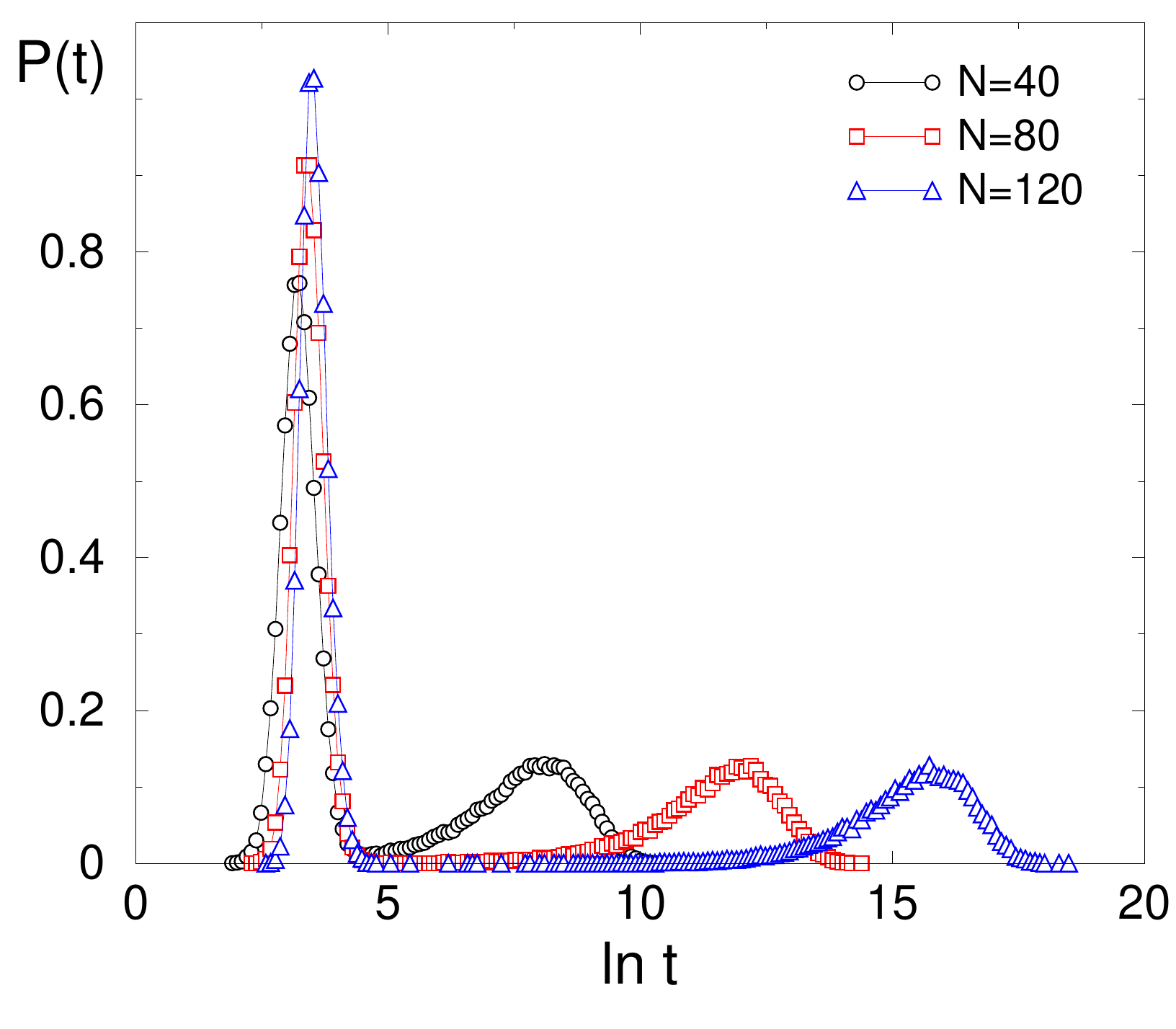}}
  \caption{The consensus-time distribution for the balanced initial condition
    when $\epsilon=0.03$ for $N=40$, 80 and 120.}
  \label{fig:tdist}
\end{figure}

Deep in the ultra-slow regime, these two disparate time scales can be readily
quantified.  For example, for $\epsilon=0.03$, the average consensus time
$\langle t \rangle$ over $10^5$ realizations is approximately
$1304, ~57104, ~2,477,250$ for $N=40, ~80, ~120$.  However, the underlying
distribution of consensus times consists of two widely separated peaks
(Fig.~\ref{fig:tdist}).  These peaks are located at roughly $t\approx 23.8$
and $3.3\times10^3$ for $N=40$, $t\approx 29$ and $1.6\times 10^5$ for
$N=80$, and $t\approx 33$ and $8.9\times 10^6$ for $N=120$.  The smaller of
these two times grows close to logarithmically with $N$, while the larger
time grows roughly exponentially with $N$.  Thus the average consensus time
is not a useful measure of how long it takes a given realization of the
population to reach consensus.

To appreciate the dynamical source of these disparate time scales for
$\epsilon<\epsilon_c$, it is useful to trace individual state-space
trajectories.  Figure~\ref{fig:traj} shows two such trajectories for
$\epsilon= 0.03$ and $N=80$.  One (magenta) corresponds to quick consensus,
in which the trajectory moves quasi-systematically from the initial state of
$(a,b)=(\frac{1}{2},\frac{1}{2})$ to consensus at (0,0) in a time of roughly
$29.5$.  The other trajectory (multiple colors) shows ultra-slow approach to
consensus.  The blue portion shows the first 100 steps, where the population
state quickly goes from $(\frac{1}{2},\frac{1}{2})$ to a metastable state
near the fixed point $\gamma_+\approx (0.968,0.032)$.  The green portion
shows the trajectory in the time range between $t\approx 100$ and $0.999\,T$,
where $T=126,456$ is the consensus time for this trajectory.  This part of
the trajectory wanders stochastically about the fixed point $\gamma_+$ until
a large fluctuation drives this trajectory outside the local basin of
attraction, after which the consensus fixed point at $(1,1)$ is reached (red
portion of the trajectory).

\begin{figure}[ht]
  \centerline{\includegraphics[width=0.35\textwidth]{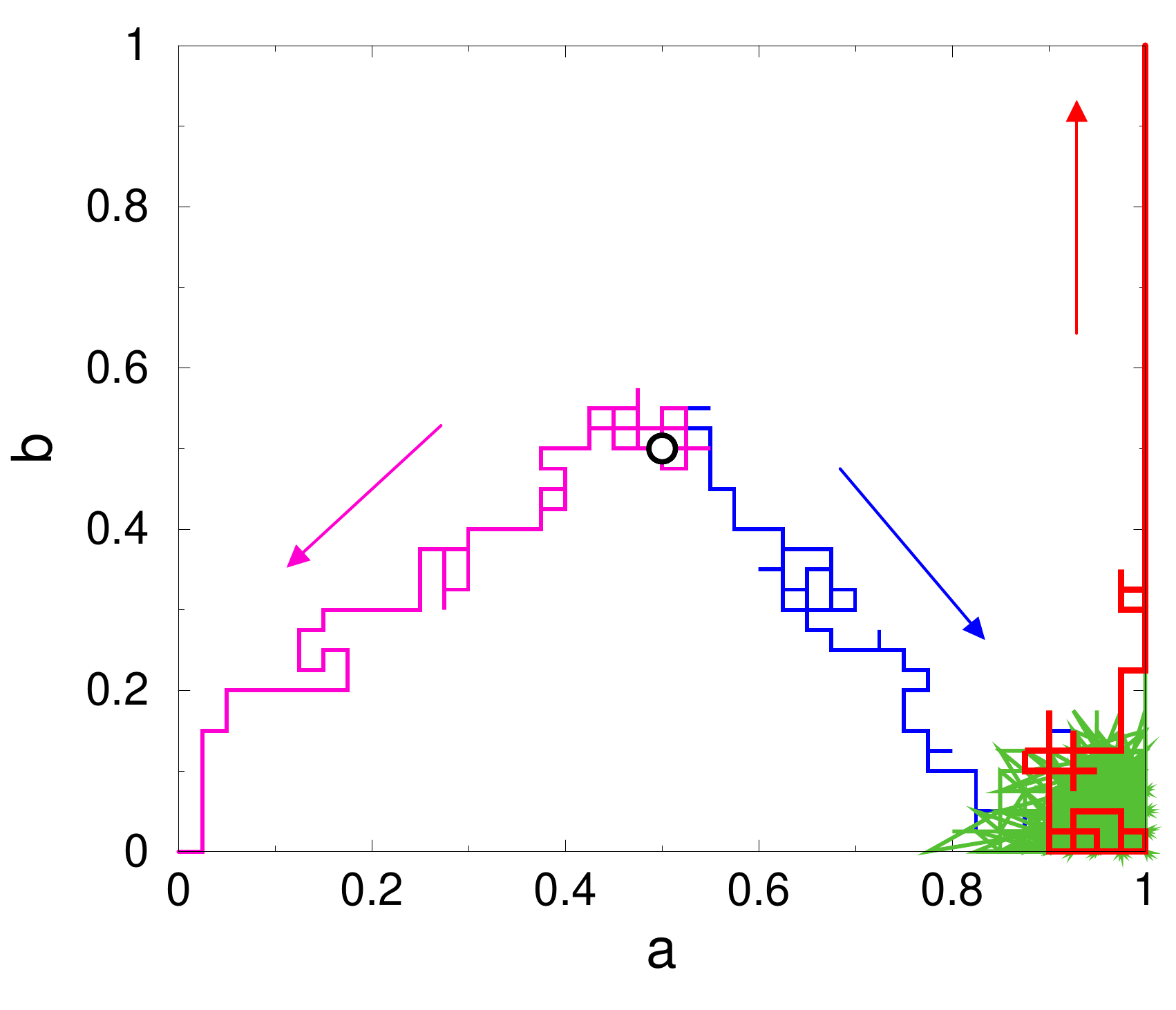}}
  \caption{Two representative state-space trajectories for $\epsilon=0.03$
    and $N=80$ that both start from $(\frac{1}{2},\frac{1}{2})$: (magenta) a
    trajectory that quickly goes to consensus at $(0,0)$; (blue/green/red) a
    trajectory that quickly reaches a metastable state near the fixed point
    $\gamma_+$ before eventually reaching consensus. }
  \label{fig:traj}
\end{figure}

When $\epsilon=\epsilon_c$, the boundary between the logarithmic and
exponential dependence of the average consensus time, we might have
anticipated`q that $\langle t\rangle$ would scale algebraically with $N$.
However, simulations indicate logarithmic scaling at $\epsilon = \epsilon_c$
(Fig.~\ref{fig:tav}(b)).  We also might have anticipated that the consensus
time is an asymptotically non-self-averaging random quantity in the
intermediate regime $\epsilon_c\leq \epsilon\leq \epsilon^c$ because of the
presence of the fixed points $\gamma_\pm$.  However, simulations suggest that
the consensus time is self-averaging for all $\epsilon \geq
\epsilon_c$. Specifically, we found that the fluctuation measure
$R\equiv \sqrt{\langle t^2\rangle}/\langle t\rangle \to 1$ faster than
algebraically in $N$ when $\epsilon>\epsilon^c$, and as $N^{-1/2}$ when
$\epsilon = \epsilon_c$.  Thus we conclude that the consensus time $t$ is a
self-averaging random  quantity that scales as $\ln N$ for all
$\epsilon \geq \epsilon_c$.

Majority rule dynamics in a two-class population favors quick consensus
(logarithmic in population size) when individuals heed members of the other
class with probability greater than 11\%.  Otherwise, a polarized state
occurs for appropriate initial conditions.  In a polarized state, one class
strongly prefers the $+1$ opinion state and the other class prefers the $-1$
state, even though neither class has an internal preference for a given
opinion.  This polarized state is practically eternal in that the escape time
to reach consensus scales exponentially with population size.  Such freezing
in of polarization seems characteristic of the current political
climate~\cite{adamic2005political,wang2017follow,wang2018great,Sokolov} and
illustrates the crucial role of interactions between different classes of
individuals in fostering either camaraderie or animosity.  Our work also
raises many interesting challenges, such as understanding homophilous
majority rule on more realistic or dynamically evolving networks,
heterogeneous groupings, and more than two opinion states.

\bigskip\noindent We thank Mauro Mobilia for helpful suggestions and
manuscript comments.  SR was partially supported by NSF grant DMR-1910736.

\bibliographystyle{apsrev4-1}

%

\pagebreak

\onecolumngrid
\begin{center}
  \textbf{\large Supplementary Material for Divergence and Consensus in Majority Rule}\\[.2cm]
  P. L. Krapivsky$^{1}$ and S. Redner$^{2}$\\[.1cm]
  {\itshape ${}^1$Department of Physics, Boston University, Boston, MA, 02215 USA\\
  ${}^2$Santa Fe Institute, 1399 Hyde Park Road, Santa Fe, NM, 87501 USA}
\end{center}

\setcounter{equation}{0}
\setcounter{figure}{0}
\setcounter{table}{0}
\setcounter{page}{1}
\renewcommand{\theequation}{S\arabic{equation}}
\renewcommand{\thefigure}{S\arabic{figure}}
\renewcommand{\bibnumfmt}[1]{[S#1]}
\renewcommand{\citenumfont}[1]{S#1}

\maketitle

\section{Rate Equations}

We derive rate equations that provide the deterministic description of the
evolution of the homophilous majority-rule (HMR) model.  We start with the
simplest case of $\epsilon=1$ when the type of the individual does not
influence the behavior.  We thus focus on the time evolution of individuals
with a given opinion. The density $\rho(t)$ of individuals with opinion $+1$
evolves according to
\begin{equation}
\label{rho-eq-sm}
\dot \rho = F(\rho), \qquad F(\rho)=\rho^2(1-\rho)-\rho(1-\rho)^2\,.
\end{equation}
The gain term account fir a group that consists of two individuals with
opinion $+1$ and one individual with opinion $-1$; the loss term accounts for
a group of the opposite composition. The time unit is chosen in such way that
each individual is potentially updated once per unit time.

For the general HMR model, we denote by $a$ and $b$ the densities of
individuals of type A and B that are in the $+1$ opinion state.  Possible
states in a group of three individuals can be obtained by expanding the
multinomial $(A_++A_-+B_++B_-)^3$, where the subscripts denote the opinion
state of a given individual.  For example, the state $A_+A_-B_+$ (with
multiplicity 6) corresponds a heterogeneous group in the opinion state $++-$.
When majority rule is applied, this group will go to the opinion state $+++$
with probability $6\epsilon a(1-a)b$.  All the other terms in the multinomial
expansion can be explained similarly.  As a result of this enumeration of all
states, the time dependence of the densities $a$ and $b$ are given by the
rate equations
\begin{equation}
\label{RE-sm}
\dot a = F(a) + \epsilon G(a,b),\qquad \dot b = F(b)+\epsilon G(b,a)\,,
\end{equation}
with $F(x)$ defined in \eqref{rho-eq-sm} and 
\begin{equation}
\label{G:def-sm}
G(x,y) = (1-x)\big[2xy+y^2\big] - x\big[2(1-x)(1-y)+(1-y)^2\big]\,.
\end{equation}

Equations \eqref{RE-sm} reduce to \eqref{rho-eq-sm} when $\epsilon=0$. In this
extreme limit of the HMR model, individuals of type A and B do not ``talk" to
each other, so an individual is potentially updated only if the two other
members of the group are of the same type. Therefore the overall rate is four
times smaller than the rate used in \eqref{rho-eq-sm} and the rate equations
would be $\dot a = \frac{1}{4}F(a)$ and $\dot b = \frac{1}{4}F(b)$ in the
units of time used in \eqref{rho-eq-sm}. The choice of time units in \eqref{RE-sm}
is made on aesthetic grounds to eliminate prefactors.  As another consistency
check, we set $\epsilon=1$ when individuals are ``blind" and react
independently on the type.  Summing Eqs.~\eqref{RE} gives $\rho=(a+b)/2$
satisfies $\dot \rho = 4F(\rho)$, with the factor of four reflecting the
different choice of time in units in \eqref{rho-eq-sm} and \eqref{RE-sm}.

\section{Fixed Points}

The flow field associated with the dynamical system defined by
Eqs.~\eqref{RE-sm} and \eqref{G:def-sm} are displayed in Fig.~\ref{fig:flow}
of the main text for representative values of $\epsilon$ in each of the three
distinct domains of behavior: (a) $\epsilon>\epsilon^c$, with
$\epsilon^c=\frac{1}{5}$, (b) $\epsilon$ between $\epsilon_c$ and
$\epsilon^c$, with $\epsilon_c=\frac{1}{9}$ and (c) $\epsilon<\epsilon_c$.
These flow fields were obtained by using the StreamPlot function in
Mathematica.

For $\epsilon> \epsilon^c$, there are three trivial fixed points that are
located at
\begin{align}
\label{trivial:FP-sm}
(0,0),\qquad \big(\tfrac{1}{2},\tfrac{1}{2}\big), \qquad
(1,1).
\end{align}
These fixed points exist for all values of $\epsilon$ in the range $[0,1]$.
The fixed points $(0,0)$ and $(1,1)$ are always stable and correspond to the
consensus states.  The symmetric fixed point $\big(\tfrac{1}{2},\tfrac{1}{2}\big)$ is a saddle node for this range of
$\epsilon$ and it corresponds to a balanced and polarized state in which half
the A's and half the B's in the population are in each opinion state.  If the
initial condition is not on the co-diagonal  $a+b=1$, the dynamics quickly drives the
population to one of the consensus states (Fig.~\ref{fig:flow}(a)).  If,
however, the initial condition lies on the co-diagonal   $a+b=1$ (with neither $a=1$
or $b=1$), rate equations predict that the population slowly approaches the symmetric fixed point
$\big(\tfrac{1}{2},\tfrac{1}{2}\big)$.

For intermediate values of $\epsilon$ that lie between $\epsilon_c$ and
$\epsilon^c$, in addition to the fixed points \eqref{trivial:FP-sm}, two new
fixed points emerge. These fixed points are symmetrically located on the
co-diagonal $a+b=1$ at
\begin{equation}
\label{gamma:+-sm}
 \gamma_\pm=\frac{1}{2}\left(1\pm \Gamma\,, 1\mp \Gamma\right), \qquad \Gamma(\epsilon)=\sqrt{\frac{1-5\epsilon}{1-\epsilon}}\,,
\end{equation}
see Fig.~\ref{fig:flow}(b).  For values of $\epsilon$ in this intermediate
range, the symmetric fixed point is now unstable and the new fixed points
$\gamma_\pm$ are saddle nodes.  If the initial densities are not on the
co-diagonal $a+b=1$, the population again quickly approaches one of the
consensus fixed points.  If the initial densities are on the co-diagonal
$a+b=1$ (except at $(\frac{1}{2},\frac{1}{2})$), then the population
approaches one of the fixed points $\gamma_\pm$.  If the initial densities
are $(\frac{1}{2},\frac{1}{2})$, the population remains at this fixed point.

Finally, for $\epsilon<\epsilon_c$, the fixed points $\gamma_\pm$ become
stable attractors and four new saddle nodes $\delta_i$ emerge ($i=1,2,3,4$)
that are symmetrically located about $a+b=1$ (Fig.~\ref{fig:flow}(c)).  We
order these fixed points by the magnitude of their vertical coordinates.
There now exists two basins of attraction, one in the SE corner and one in
the NW corner of the phase space $\{(a,b)\in [0,1]^2\}$ of the system for the
fixed points $\gamma_\pm$ (Fig.~\ref{fig:flow}(d)).  Points outside these
basins are attracted to one of the consensus fixed points.  The basin
boundaries are the separatrices from $\big(\tfrac{1}{2},\tfrac{1}{2}\big)$ to
each $\delta_i$ and also from four specific points on the boundary of the
phase space to $\delta_1$.  Points on these separatrices are attracted to one
of the $\delta_i$, apart from the symmetric fixed point
$\big(\tfrac{1}{2},\tfrac{1}{2}\big)$.  When $\epsilon\to 0$, the stable
fixed points $\gamma_\pm$ approach (0,1) and (1,0), and the saddle nodes
$\delta_i$ approach $(\frac{1}{2},0)$, $(0,\frac{1}{2})$, $(1,\frac{1}{2})$,
and $(\frac{1}{2},1)$.  In the case of $\epsilon=0$, the full system breaks
into two independent subpopulations that consists of only A and only B that
each undergo pure majority-rule dynamics.

The location of the fixed points \eqref{trivial:FP-sm} are obvious and the
locations of the fixed points $\gamma_\pm$ can be determined by hand without
too much difficulty.  The determination of the locations of the fixed points
$\delta_i$ are complicated and we use Mathematica for the computations that
follow.  The symmetry of the problem allows us to express locations of these
fixed points $\delta_i=(x_i,y_i)$ through the coordinates of one fixed point:
\begin{equation}
\delta_1 = (x_1,y_1), \quad \delta_2=(1-y_1,1-x_1), \quad \delta_3=(y_1,x_1), \quad \delta_4=(1-x_1,1-y_1)
\end{equation}
We have chosen [Fig.~\ref{fig:flow}(c)] as $\delta_1$ the fixed point with
the smallest vertical coordinate, and ordered the other fixed points by the
magnitude of the vertical component. The expressions for $x_1, y_1$ are
rather cumbersome and we find
\begin{equation}
\label{x1-sm}
2x_1=1 + \sqrt{1-\frac{1+3\epsilon}{2}\left[\frac{1-3\epsilon}{1-\epsilon}+\sqrt{\frac{1-9\epsilon}{1-\epsilon}}\right]}
\end{equation}
The dependences of the fixed point locations on $\epsilon$ and their
stability as a function of $\epsilon$ are shown in Fig.~\ref{fig:fp} in the main text.

To determine the stability of the fixed points, we need to evaluate the Jacobian matrix
\begin{align}
  J &= 
      \left(
\begin{array}{cc}
{\displaystyle \frac{\partial [F(a) + \epsilon G(a,b)]}{\partial a}} & {\displaystyle \frac{\partial [F(a) + \epsilon G(a,b)]}{\partial b}} \\[4mm]
{\displaystyle \frac{\partial [F(b)+\epsilon G(b,a)]}{\partial a}} & {\displaystyle\frac{\partial [F(b)+\epsilon G(b,a)]}{\partial b}}\\
\end{array}
\right)~.
\end{align}
Using the definitions \eqref{rho-eq-sm} and \eqref{G:def-sm} for $F$ and  $G$, we have
\begin{align*}
\frac{\partial [F(a) + \epsilon G(a,b)]}{\partial a} &= 6a - 6a^2 -1+ 
 \epsilon \big[4a + 6b - 8ab-2b^2 -3 \big],\\[2mm]
\frac{\partial [F(a) + \epsilon G(a,b)]}{\partial b} &=2\epsilon \big[a(2-a-b) + (1-a) (a+b)\big], \\[2mm]
\frac{\partial [F(b) + \epsilon G(b,a)]}{\partial a} &=2\epsilon \big[b(2-a-b) + (1-b) (a+b)\big],\\[2mm]
\frac{\partial [F(b) + \epsilon G(b,a)]}{\partial b} &= 6b - 6b^2 -1 + 
 \epsilon \big[4b + 6a - 8ab-2a^2 -3\big]\,,
\end{align*}  
at each fixed point: The eigenvalues of the Jacobian evaluated at each fixed
point determine its stability.

\begin{figure}[ht]
  \centerline{\subfigure[]{\includegraphics[width=0.32\textwidth]{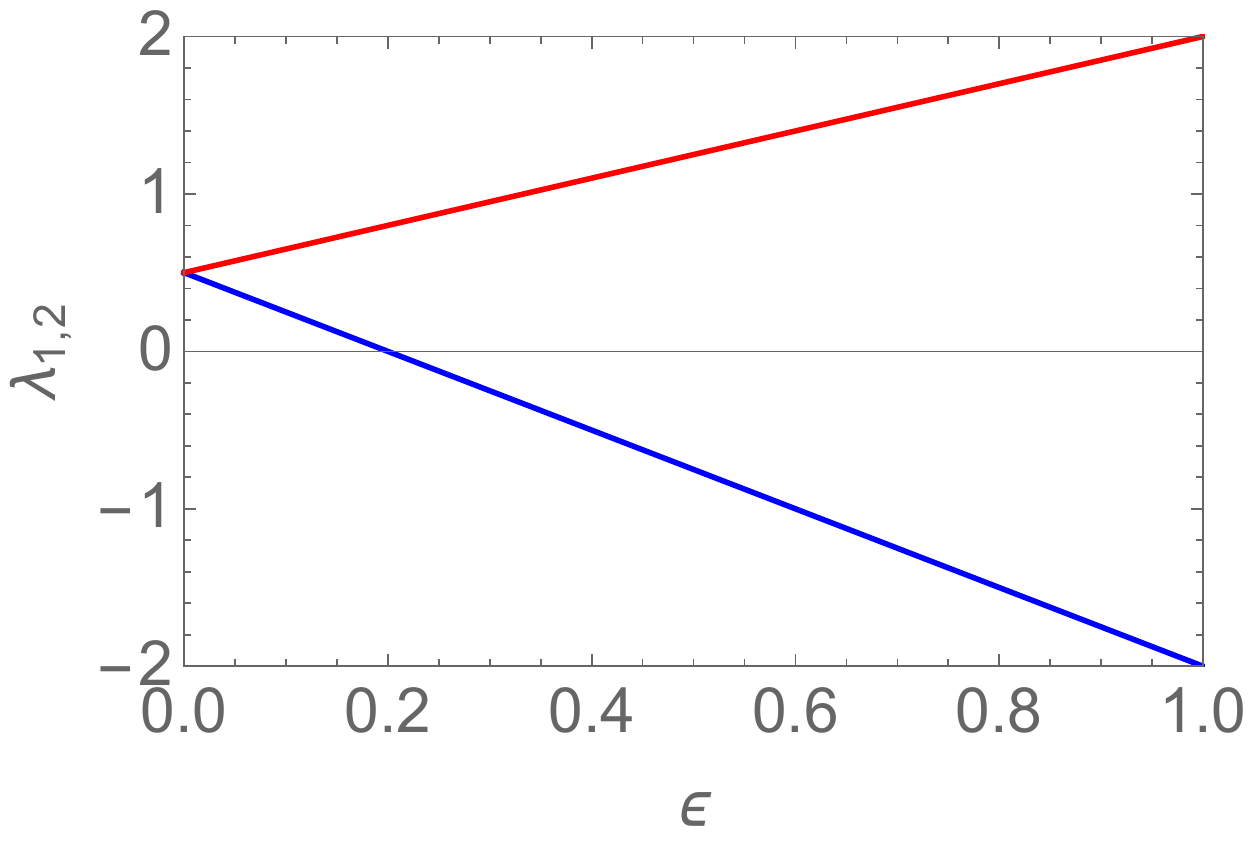}}\quad
    \subfigure[]{\includegraphics[width=0.32\textwidth]{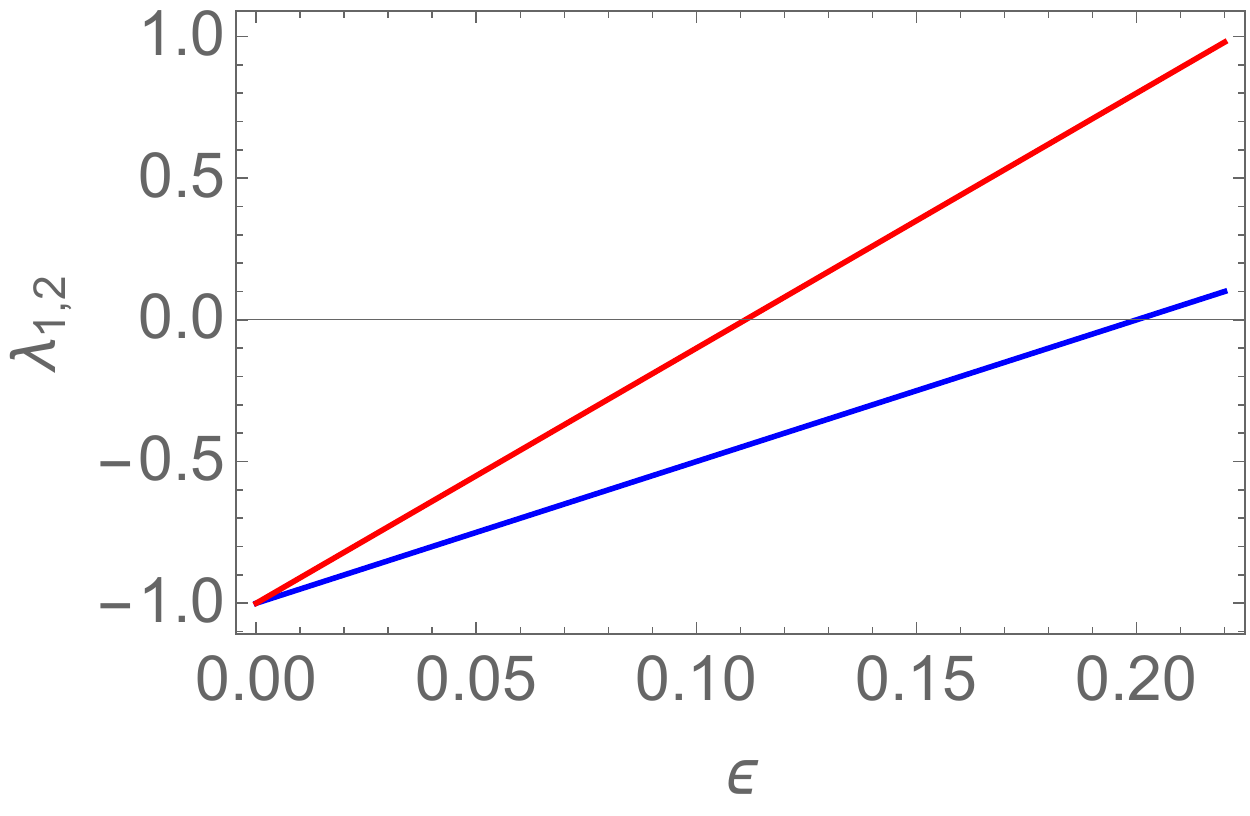}}\quad
    \subfigure[]{\includegraphics[width=0.34\textwidth]{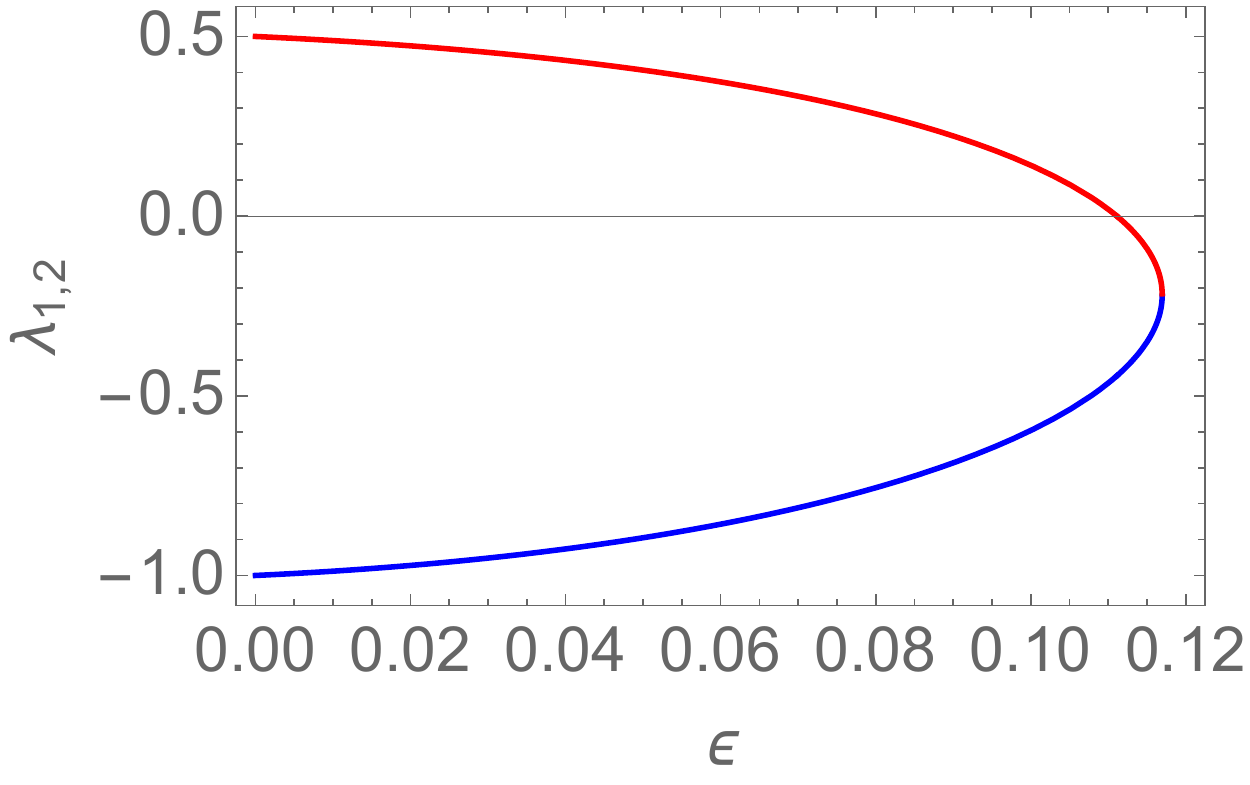} }}
  \caption{The $\epsilon$ dependence of the eigenvalues of the Jacobian
    matrix at: (a) the symmetric fixed point
    $\big(\tfrac{1}{2},\tfrac{1}{2}\big)$ in the range
    $0\leq \epsilon\leq 1$; (b) the fixed points $\gamma_\pm$ in the range
    $\epsilon\leq \epsilon^c=\frac{1}{5}$ where these fixed points exist; (c)
    the fixed points $\delta_i$ in the range
    $\epsilon\leq \epsilon_c=\frac{1}{9}$ where these fixed points exist.}
\label{fig:lambda}
\end{figure}

At the consensus fixed points, the Jacobian is diagonal,
$J=\text{diag}[-1-3\epsilon, -1-3\epsilon]$, so the eigenvalues are negative,
$\lambda_1=\lambda_2=-1-3\epsilon$, and both consensus fixed points are
stable.  At the symmetric fixed point $\big(\tfrac{1}{2},\tfrac{1}{2}\big)$,
the Jacobian matrix and its corresponding eigenvalues are
\begin{align}
  J &= 
      \left(
\begin{array}{cc}
{\frac{1}{2}\displaystyle (1-\epsilon) } & {\displaystyle 2\epsilon} \\[4mm]
{\displaystyle 2\epsilon} & {\frac{1}{2}\displaystyle  (1-\epsilon) }
\end{array}
\right)\,,
  \qquad\qquad \lambda_1= \tfrac{1}{2}(1+3\epsilon)\,, \qquad \lambda_2= \tfrac{1}{2}(1-5\epsilon)\,.
\end{align}
These eigenvalues [Fig.~\ref{fig:lambda}(a)] vary linearly with $\epsilon$.
One eigenvalue is always positive and the second is positive when
$\epsilon<\epsilon^c=\frac{1}{5}$, so that the symmetric fixed point is
unstable in this range.  The second eigenvalue is negative when
$\epsilon^c<\epsilon$, which implies that the symmetric fixed point is a
saddle node in this range.

At the reflection-symmetric fixed points $\gamma_\pm$, the Jacobian matrix
and its corresponding eigenvalues are 
\begin{align}
  J &= 
      \left(
\begin{array}{cc}
{\displaystyle -1+7\epsilon} & {\displaystyle 2\epsilon} \\[4mm]
{\displaystyle 2\epsilon} & {\displaystyle-1+7\epsilon}
\end{array}
\right)\,,\qquad\qquad  \lambda_1 = -1+9\epsilon\,,\qquad \lambda_2=-1+5\epsilon \,.
\end{align}
The dependence of these eigenvalues on $\epsilon$ is shown in
Fig.~\ref{fig:lambda}(b).  The displayed range of the mixing parameter is 
$\epsilon<\epsilon^c=\frac{1}{5}$, where the fixed points $\gamma_\pm$
actually exist.  One eigenvalue is positive and the other is negative for
$\epsilon>\epsilon_c$, which means that these two fixed points are saddle
nodes in this range.  For $\epsilon<\epsilon_c=\frac{1}{9}$, both eigenvalues
are negative, so that the fixed points $\gamma_\pm$ are stable nodes in this range.
This change in behavior is reflected in the qualitative change in the flow
field between Figs.~\ref{fig:flow}(b) and \ref{fig:flow}(c).

At the four non-symmetric fixed points $\delta_i$, the expressions for the
associated eigenvalues are extremely cumbersome.  The Mathematica expressions
for these eigenvalues and related information about the rate equation
solution are available on the SR's publications webpage
(\url{http://tuvalu.santafe.edu/~redner/pubs.html}).  The essential feature
is that the eigenvalue $\lambda_1$ is positive and $\lambda_2$ is negative in
the entire range $\epsilon<\epsilon_c$.  The relevant range for this figure
is $\epsilon< \epsilon_c=\frac{1}{9}$, where the fixed points $\delta_i$
actually exist.  Thus each of the fixed points $\delta_i$ is a saddle node
for all $0<\epsilon<\epsilon_c$.

\section{Fate of the balanced system for $\epsilon<\epsilon_c$}

The balanced state $\big(\tfrac{1}{2},\tfrac{1}{2}\big)$ is a fixed point of
the deterministic rate equations, but not of the actual stochastic HMR model.
When $\epsilon\geq \epsilon_c$, the HMR model with the initial condition
$(a,b)=\big(\tfrac{1}{2},\tfrac{1}{2}\big)$ quickly reaches one of the two
consensus states.  When $\epsilon < \epsilon_c$, however, fluctuations may
drive the system into the basin of attraction of one of the stable fixed
points $\gamma_\pm$.  Indeed, when the population size is sufficiently large,
the dynamics will be effectively deterministic, which allows us to make a
simple estimate for the probability to reach the fixed point $\gamma_+$.

While we do not know the exact location of the separatrices analytically,
Fig.~\ref{fig:flow}(c) shows that the separatrices lie inside the wedge
formed by the two rays that emanate from
$\big(\tfrac{1}{2},\tfrac{1}{2}\big)$ to $\delta_1$ and to $\delta_2$. The
angle $\theta$ between these rays therefore equals the angle between vectors
$\big(x_1-\frac{1}{2},y_1-\frac{1}{2}\big)$ and
$\big(\frac{1}{2}-y_1,\frac{1}{2}-x_1\big)$.  Thus an upper bound for
probability $\Pi(\epsilon)$ that the state of the system enters  the
basin of attraction of $\gamma_+$ is
\begin{equation}
\label{prob-sm}
\Pi(\epsilon) \leq  \frac{\theta}{\pi} = \frac{1}{\pi}\cos^{-1}\!\left[\frac{2\big(x_1-\frac{1}{2}\big)\big(\frac{1}{2}-y_1\big)}{\big(x_1-\frac{1}{2}\big)^2+\big(\frac{1}{2}-y_1\big)^2}\right]\,.
\end{equation}
After substituting the cumbersome expression for $x_1$ and $y_1$ from
Eq.~\eqref{x1-sm} into \eqref{prob-sm}, we obtain a rather simple expression for
the upper bound:
\begin{equation}
\label{prob-simple-sm}
\Pi(\epsilon) \leq  \frac{1}{\pi}\cos^{-1}\!\left[\frac{8\epsilon}{1-2\epsilon+9\epsilon^2}\right]\,.
\end{equation}

\begin{figure}[ht]
  \includegraphics[width=0.35\textwidth]{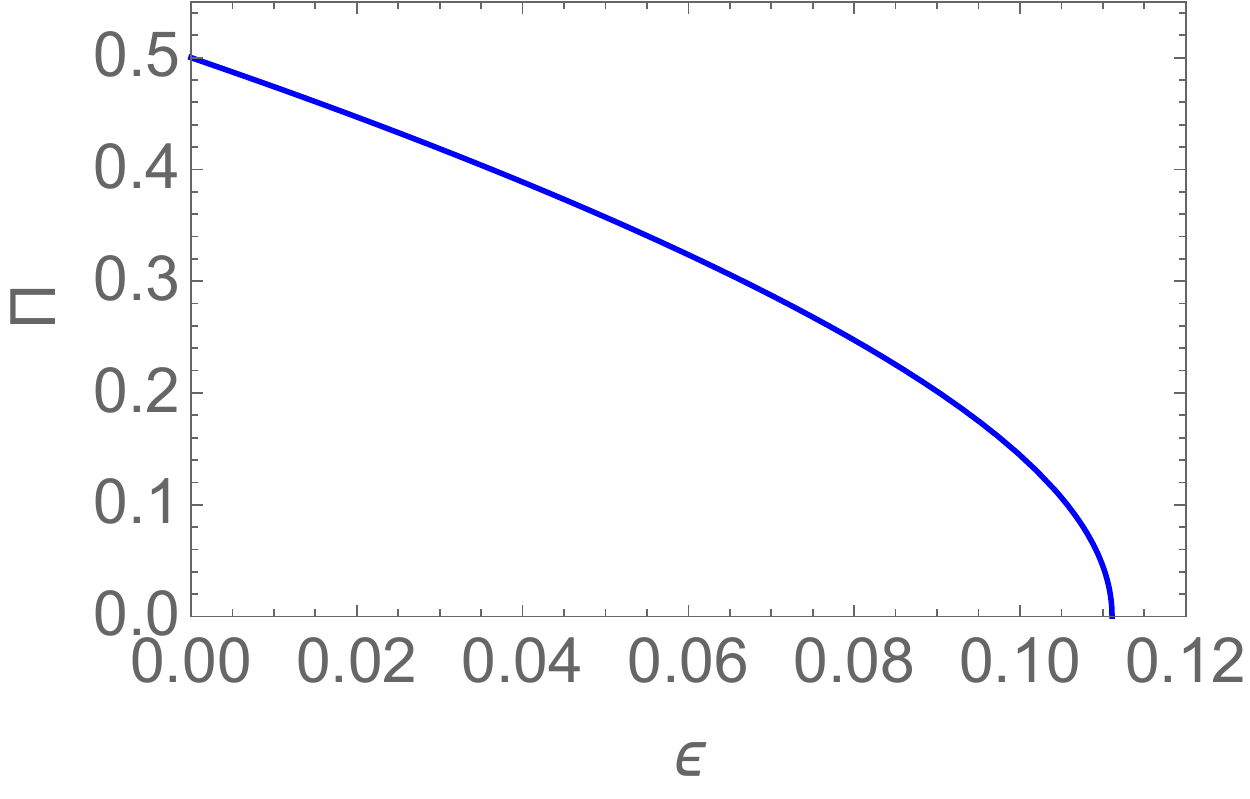}
  \caption{The $\epsilon$ dependence of Eq.~\eqref{prob-simple-sm} for
    $\Pi(\epsilon)$, the probability for the system to enter the basin of
    attraction of the fixed point $\gamma_+$. }
\label{fig:Prob-Pi}
\end{figure}

The behavior of this bound for the probability $\Pi(\epsilon)$ is shown in
Fig.~\ref{fig:Prob-Pi}.  In the limiting case of $\epsilon=0$, the upper
bound \eqref{prob-simple-sm} gives $\frac{1}{2}$ which is exact.  Indeed, for
this case, individuals of type A and B do not interact, so that there are
four possible consensus states $(0,0), ~(1,0),~(0,1),~(1,1)$, each of which
are reached equiprobably when the starting state is
$(a,b)=\big(\tfrac{1}{2},\tfrac{1}{2}\big)$.  Thus the probability $\Pi(0)$
to reach the $(1,0)$ or the $(0,1)$ consensus state is $\frac{1}{2}$.

\section{Exponential HMR Model}

According to the update rules of the HMR, the opinion change event $A_+\to A_-$
occurs in the following situations:
\begin{enumerate}
\item $A_+ A_-A_-\longrightarrow A_- A_-A_-$\qquad  rate 1;
\item $A_+ A_-B_-\longrightarrow A_- A_-B_-$\qquad  rate $\epsilon$;
\item $A_+ B_-B_-\longrightarrow A_- B_-B_-$\qquad  rate $\epsilon$.
\end{enumerate}
For the latter two cases, the opinion change $A_+\to A_-$ occurs at the same
rate $\epsilon$ when the group opinion state is $--+$, as long as at least
one of the individuals in the group is a B.

A natural alternative is that a person is motivated to change opinion by an
individual from the same class with weight 1 and by an individual from the
opposite class with weight $\epsilon$. 
This motivates the following modification of majority rule:
\begin{enumerate}
\item $A_+ A_-A_-\longrightarrow A_- A_-A_-$\qquad  rate 1;
\item $A_+ A_-B_-\longrightarrow A_- A_-B_-$\qquad  rate $\epsilon$;
\item $A_+ B_-B_-\longrightarrow A_- B_-B_-$\qquad  rate $\epsilon^2$.
\end{enumerate}
The qualitative behaviors of the original and this exponential version of HMR
are essentially identical, so in the main text we investigated only the
original version.  For completeness, we outline the dynamical behavior of
this exponential HMR.

The rate equations that govern the evolution of this exponential HMR model are
\begin{equation}
\label{RE-H-sm}
\dot a = F(a) + H(a,b),\qquad \dot b = F(b)+H(b,a),
\end{equation}
with the same $F(x)$ as in \eqref{rho-eq-sm}, and
\begin{equation}
\label{H:def-sm}
H(x,y) = 2\epsilon x(1-x)(2y-1) + \epsilon^2\big[(1-x)y^2 -x(1-y)^2\big]\,.
\end{equation}
The critical values of the mixing parameter are now
\begin{equation}
\epsilon^c = \frac{1}{3}\,, \qquad \epsilon_c = \sqrt{5}-2
\end{equation}
In addition to the trivial fixed points \eqref{trivial:FP-sm}, two additional
fixed points $\gamma_\pm$ that are located at
\begin{equation}
\label{gamma-new-sm}
 \gamma_\pm=\tfrac{1}{2}\left(1\pm \Gamma\,, 1\mp \Gamma\right), \qquad \Gamma=\frac{\sqrt{(1+\epsilon)(1-3\epsilon)}}{1-\epsilon}\,,
\end{equation}
emerge from $\beta$ when $\epsilon<\epsilon^c$.  Four additional fixed points
emerge from $\gamma_\pm$ at $\epsilon=\epsilon_c$ and the analog of
\eqref{x1-sm} is
\begin{equation}
\label{x1-new-sm}
2x_1=1 + \frac{\sqrt{1-\epsilon+3\epsilon^2+ \epsilon^3 -(1+\epsilon)\sqrt{(1-\epsilon^2)(1-4\epsilon-\epsilon^2)}}}{\sqrt{2(1-\epsilon)}}~.
\end{equation}
In the $\epsilon\to 0$ limit, the density of A's with opinion $-1$ in the
polarized state $\gamma_+$ vanishes in a different manner in two versions:
\begin{equation}
a =
\begin{cases}
  \epsilon+2\epsilon^2+5\epsilon^3+ \ldots  & \text{(HMR model)}\,,\\
  \epsilon^2+2\epsilon^3+\ldots & \text{(exponential HMR)}\,.
\end{cases}
\end{equation}
The primary new aspect of the exponential HMR model is that the polarized state
becomes more extreme in character as $\epsilon\to 0$.

Finally, we note that in the $N\to\infty$ limit, the exponential HMR model
can be interpreted as the standard MR model on a random graph that admits a
partitioning into two complete sub-graphs of size $N$.  Each subgraph
represents one of the classes, either A and B, and any two nodes from
different classes are connected with probability $\epsilon$.

\end{document}